\documentclass[a4paper]{jpconf}
\usepackage{graphicx}
\usepackage{hyperref}
\usepackage{amsmath,amssymb}
\DeclareMathOperator{\sech}{sech}

\newcommand{\bfE}{\mathbf{E}}
\newcommand{\bfB}{\mathbf{B}}

\newcommand{\bfu}{\mathbf{u}}
\newcommand{\bfxhat}{\mathbf{\hat{x}}}
\newcommand{\bfzhat}{\mathbf{\hat{z}}}
\newcommand{\bfJ}{\mathbf{J}}

\begin{document}
\title{Where should MMS look for electron diffusion regions?}

\author{G. Lapenta$^1$, M. Goldman$^2$, D. Newman$^2$, S. Markidis$^3$}
\address{$^1$ Center for mathematical Plasma Astrophysics, University of Leuven, KU Leuven, Belgium}
\address{$^2$ Department of Physics, University of Colorado, Boulder, USA}
\address{$^3$ HPCVIZ, KTH, Stockholm, Sweden}

\ead{giovanni.lapenta@wis.kuleuven.be}

\begin{abstract}
A great possible achievement for the MMS mission would be crossing electron diffusion regions (EDR). EDR are regions in proximity of reconnection sites where electrons decouple from field lines, breaking the frozen in condition. Decades of research on reconnection have produced a widely shared map of where EDRs are. 
We expect reconnection to take place around a so called x-point formed by the intersection of the separatrices dividing inflowing from outflowing plasma.  The EDR forms around this x-point as a small electron scale box nested inside a larger ion diffusion region. 
But this point of view is based on a 2D mentality. We have recently proposed that once the problem is considered in full 3D,  secondary reconnection events can form [Lapenta et al., Nature Physics,  11, 690, 2015] in the outflow regions even far downstream from the primary reconnection site. 
We revisit here this new idea confirming that even using additional indicators of reconnection and even considering longer periods and wider distances the conclusion remains true:  secondary reconnection sites form downstream of  a reconnection outflow causing a sort of chain reaction of cascading reconnection sites.
If we are right, MMS will have an interesting journey even when not crossing necessarily the primary site. The chances are greatly increased that even if missing a primary site during an orbit, MMS could stumble instead on one of these secondary sites. \end{abstract}

\section{Introduction}
The Magnetospheric Multiscale (MMS) mission launched successfully on March 12, 2015 at 10:44pm. At the time of writing MMS data is  already arriving and being analysed, revolutionising our field with a quantum jump in quality and resolution.  
The goals of the mission as stated on its web site aim squarely at the core of  reconnection physics: {\it
    What determines when reconnection starts and how fast it proceeds?
    What is the structure of the diffusion region?
    How do the plasmas and magnetic fields disconnect and reconnect in the diffusion regions?
    What role do the electrons play in facilitating reconnection?
    What is the role of turbulence in the reconnection process?
    How does reconnection lead to the acceleration of particles to high energies?} [From: \url{mms.gsfc.nasa.gov}].

\begin{figure}
  \centering
  \includegraphics[width=\columnwidth]{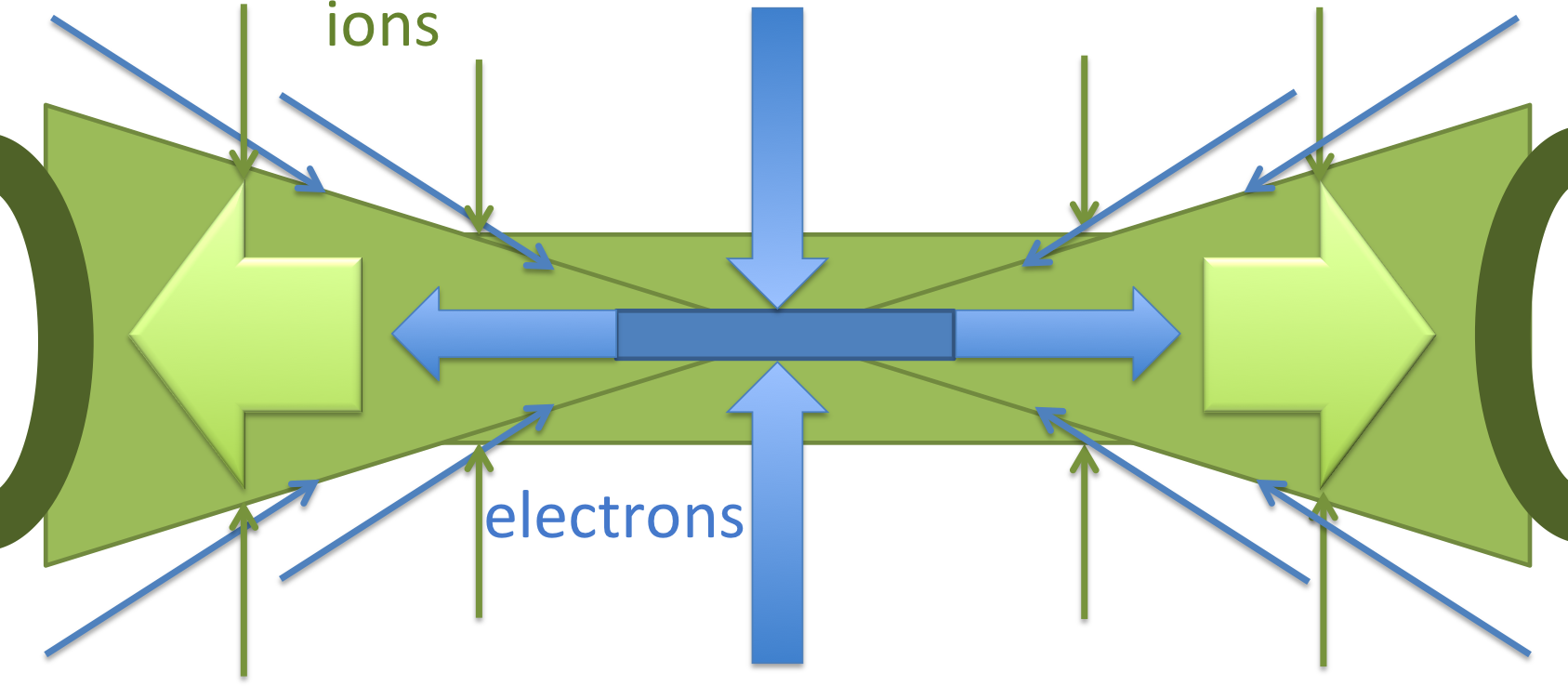}
  \caption{Cartoon of the  electron and ion flow patterns in kinetic reconnection. Only a minority of electrons and ions pass near the x-point. Around this point, an electron diffusion region forms with electrons arriving from the inflow region and being ejected as two narrow jets into the outflow (large blue arrows). The vast majority of electrons (blue arrows) and ions (green arrows) pass through the separatrices where their flow speed receives a large increase over a narrow transition layer. The area within the separatrix (in green) is filled with reconnected downstream plasma and the electrons and ions cross at the separatrices. At the exit of the reconnection region the outflow meets its environment forming two fronts (darker green). }
\label{cartoon}
\end{figure}

To answer these questions MMS must first find a reconnection site and pass through a region where the processes discussed above are happening. A typical configuration of a reconnection site is shown in Fig.~\ref{cartoon}. Reconnection has a central fulcrum, called x-point where field lines advect to from above and below, break and reconnect to eventually advect outwards left and right~\cite{biskamp,birn-priest}. The electrons an ions follow this process as well but become separated in the course of reconnection. The ions separate first in the so-called ion diffusion region (green area) where their speed separates from that of the field lines and of the electrons that remain frozen in. The electrons eventually also break their link with the field lines in the electron diffusion region (EDR, in blue). The electrons emerge from the electron diffusion region forming jet that can be deflected by the presence of guide fields~\cite{goldman2011jet}. The separatrices play a critical role as most electrons and ions do not move into the reconnecton exhaust passing via the central diffusion regions, but rather cross via the separatrices \cite{lapenta2014separatrices}. 

The central EDR is small, of the electron skin depth scale~\cite{birn-priest}\footnote{See expecially the chapter by Shay and Drake}.  A crowning achievement for MMS would be to cross a few of these regions and observe their inner structure. Where should we look for them? Where will MMS have the best chances of finding one? Based on Fig.~\ref{cartoon} the EDR should be located around the x-point where the separatrices converge.   

Figure~\ref{cartoon}  is a cartoon based on 2D understanding of reconnection. This picture is still valid in 3D only if sufficient symmetry is assumed in the out of page direction for the x-point to become what is defined as a x-line. This concept assumes that the conditions around the Earth are locally relatively azimuthal symmetric, that is the conditions do not change rapidly in the dawn-dusk direction going around the Earth on the ecliptic plane.

However, we know from satellite data that if the large scale structure of the Earth magnetosphere has some regions of slow variation in the azimuthal direction, there are in fact processes that introduce rapid and local changes in the azimuthal direction. Recently we have proposed that one among these instabilities can significantly alter our understanding of where EDR are likely to be found~\cite{lapenta2015secondary}:   drift-interchange modes downstream of regions of reconnection outflow can lead to secondary reconnection site.

The region where the outflow from reconnection come into contact with the pre-existing plasma environment forms  fronts~\cite{sitnov2009dipolarization}. The snow plower analogy is often used to describe the process: like a snow plower pushing the show to the edge of a road forms two crests of snow piling up higher than further afield, similarly the outflow from reconnection tends to form a piling up of vertical magnetic field and of plasma species. We need to recall that the outflow is generated by the Lorentz force of the curved field lines formed by reconnection~\cite{goldman2011jet}. The Maxwell stress pushes the lines to try to straighten them. This force accelerates the plasma outward, left and right of the central x-line, against the environment. This motion  encounters  resistance from the ambient plasma:  the divergence of the pressure tensor balances that of the Maxwell stress tensor. At the front we essentially have magnetic tension pushing against pressure. The force caused by the Maxwell stress tensor is equal and opposite to the divergence of the pressure tensor. 

These are the perfect conditions to give rise to the type of instabilities foreseen by Lord Rayleigh and Sir Geoffrey Ingram Taylor: the curvature of the magnetic field lines is co-directed with the pressure gradient, the conditions are met for the interchange mode~\cite{kruskal1954some,nakamura2002interchange}. However, in plasma physics,  interchange modes are much more complex and diversified than in regular fluids \cite{pritchett2010kinetic}. In the conditions typical of the magnetosphere, a fluid treatment is not sufficient and we need to seek a fully kinetic answer. The answer in this case is that the interchange mode acquires a drift nature with the growth rate better described by the standard theory of the lower hybrid drift instability that by fluid models~\cite{divin2015evolution,divin2015lower}. The dual nature of interchange and drift mode of the present instability is not unique of these conditions and has been observed in tokamak experimental investigations~\cite{poli2006experimental} in regions where unfavourable curvature plays a significant role in determining the properties of drift modes. We refer to this instability with the same nomenclature used in fusion research: {\it drift-interchange}. 

We then have a kinetic variant of the Rayleigh-Taylor (RT) mode developing along the advancing fronts. The front will  become rippled~\cite{guzdar2010simple,vapirev2013formation,lapenta2011self} with the dominant wavelength determined by kinetic theory~\cite{vapirev2013formation,divin2015evolution}. The ripples grow and tend to merge in a inverse cascade that produces fewer and bigger finger-like structures characteristic also of common fluids~\cite{2011EL.....9365001L,vapirev2013formation}.

The question we want to answer here is what happens in the region affected by these fingers?

\begin{figure}[h]
\includegraphics[width=\columnwidth]{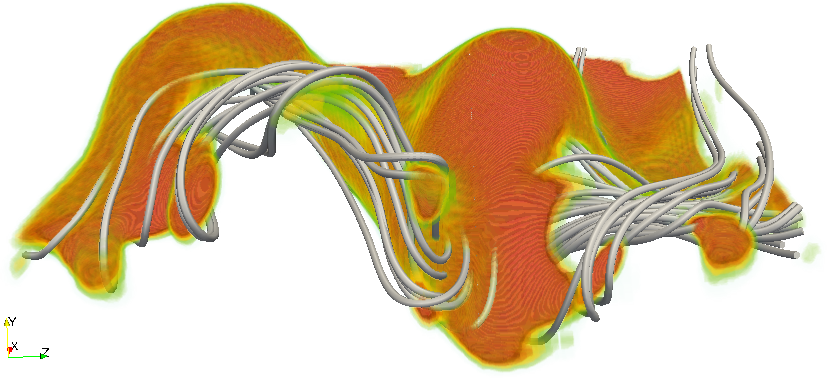}
\caption{A drift-interchange unstable front generated by the interaction of a reconnection outflow from a reconnection site. The magnetic field lines are shown in grey and are intertwined with the ripples in the front.}
\label{bello}
\end{figure}

Figure \ref{bello} shows one front with a fully developed system of RT-type ripples. The magnetic field is intertwined with the ripples and the electrons remain frozen in in the large fields of the fronts. The interaction among the ripples become then topological relevant and can lead to secondary reconnection sites. Where two interacting ripples happen to push together oppositely directed fields, a  new reconnection site can be formed.

In the remainder of the present work we will show evidence for the formation and interaction of the ripples. We will show that in some regions nulls form where field lines of opposite polarity are pushed together. We will prove that secondary reconnection develops there and we will show that under common definitions of electron slippage and breaking of the frozen in condition~cite{marty-review}, secondary EDR are formed. The results presented here are new but based on the same simulations reported in Ref.~\cite{lapenta2015secondary}. We consider here longer times where the instabilities have further developed and expand the list of   diagnostics used to make a positive identification of reconnection and of EDR.

\section{Drift-interchange destabilization of  fronts formed by reconnection outflows}
We consider an initial Harris equilibrium with density: $$n=n_0 \sech^2(2y/d_i)+n_0/10$$ in a  domain of $L_x=40d_i$, $L_y=15d_i$ and $L_z=10d_i$ with $d_i$ defined by $n_0$. The initial field includes the Harris reversal plus a guide field: $$\bfB =B_0 \bfxhat + B_0\bfzhat /10$$

\begin{figure}[h]
\hskip .15\columnwidth a) Ion Density \hskip .4\columnwidth b) $B$\\
\includegraphics[width=\columnwidth]{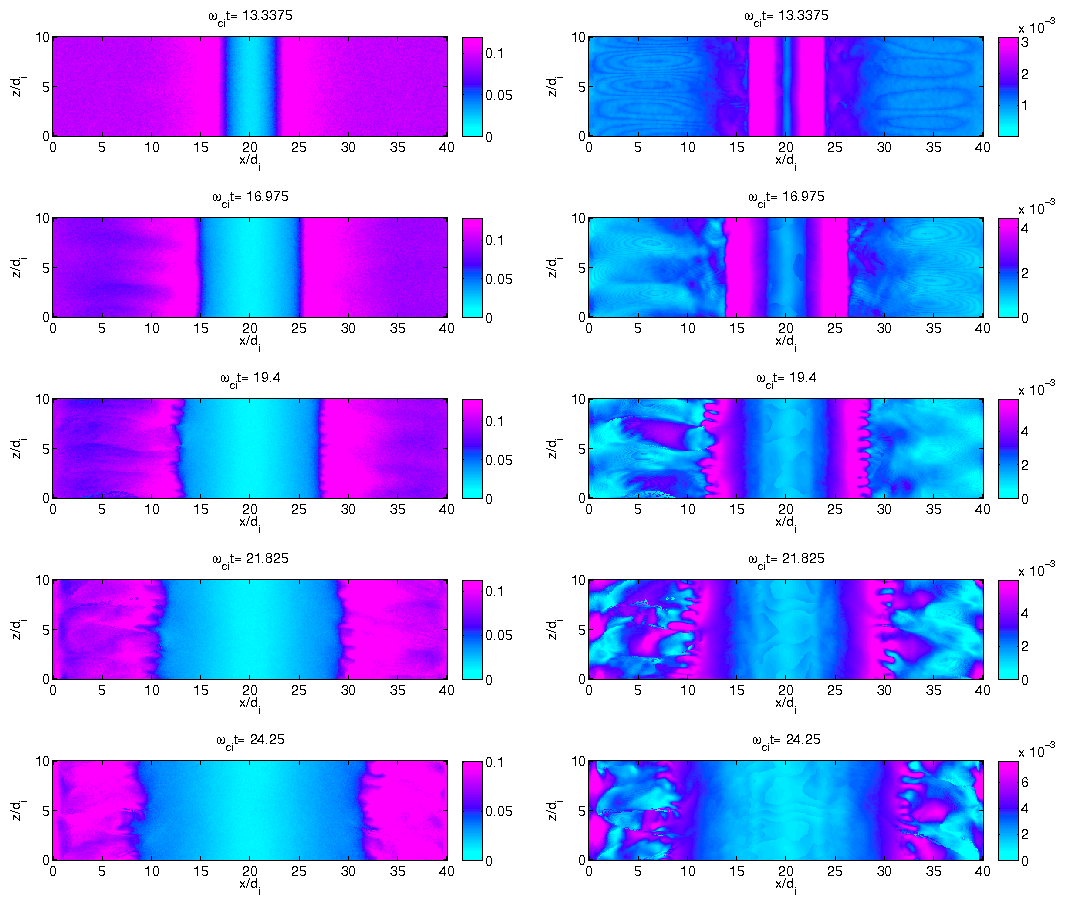}
\caption{False color representation of the density (left) and magnetic field intensity (right) on the neutral plane (defined as the plane where $|B_x|$ reaches its minimum). Different subsequent times are shown marked by the labels. All quantities are in code units.}
\label{np_den}
\end{figure}

Reconnection is initiated uniformly along the $z$ direction via a localized perturbation exponentially decreasing away from the central x-line positioned at $x=L_x/2, y=L_y/2$~\cite{Lapenta2010} and is followed using the iPic3D code~\cite{iPic3D}. Open boundary conditions are used in $x$ and $y$ to allow reconnection to progress for long times without interference from the boundary. Plasma is allowed to enter from an outside infinite reservoir along the $y$ direction and to exit from the $x$ direction. But the $z$ direction is assumed periodic.

Figure \ref{np_den} shows the evolution of the ion density and magnetic field on the neutral plane defined by the condition that $B_x$ is minimum. At each $x,z$ coordinate we locate the position $y$ where the amplitude of the magnetic field component $|B_x|$ reaches its minimum and project the false colour representation of the density and magnetic field intensity on this plane to the $xz$-plane of Fig.~\ref{np_den}. The neutral plane is initially located at $y=L_y/2$ but changes its position as the instabilities establish initially and develops in response to the changes induced by reconnection. Following the evolution on the neutral plane has the advantage of providing a 2D view of the most critical 3D plane, simplifying the presentation on a 2D page. 

Figure \ref{np_den} shows the formation from the initial straight x-line in the center of two fronts where the magnetic field piles up (right panels). Following the panels from top to bottom, we can follow the evolution of the fronts in time. Ahead of the fronts, a snow plow effect accumulates plasma density (left panels). As time progresses, the fronts move away from the central line. An instability is seen forming at the fronts leading to their rippling. The instability at the fronts extends also downstream and the plasma between the fronts and the edge of the box is perturbed. The front instability affects then not just the front proper but a vast region downstream of the fronts. It is in this vast region and not just at the fronts where we expect to see secondary reconnection.

The nature of this instability is that of a drift mode~\cite{vapirev2013formation,divin2015evolution}: the density at the front has a strong gradient in a direction normal to the field, the typical condition for drift modes~\cite{divin2015evolution}. We reported elsewhere the analysis of the mode spatial and temporal spectrum~\cite{lapenta2015secondary}. The temporal fluctuations at the front peak at the lower hybrid frequency $\omega_{lh}$ and the  wavenumber of the perturbation along $z$ is in the $(\rho_e\rho_i)^{1/2}$ range. The instability is then reasonably described by the linear theory of the lower hybrid drift instability~\cite{divin2015evolution}.  Further, theoretical investigation will be required to fully ascertain the role of curvature in these drift-interchange modes~\cite{poli2006experimental}.

\section{Energy exchanges associated with drift-interchange unstable fronts}

\begin{figure}[h]
\hskip .2\columnwidth a) $\bfJ \cdot \bfE$ \hskip .4\columnwidth b) $\bfJ \cdot \bfE^\prime$\\
\includegraphics[width=\columnwidth]{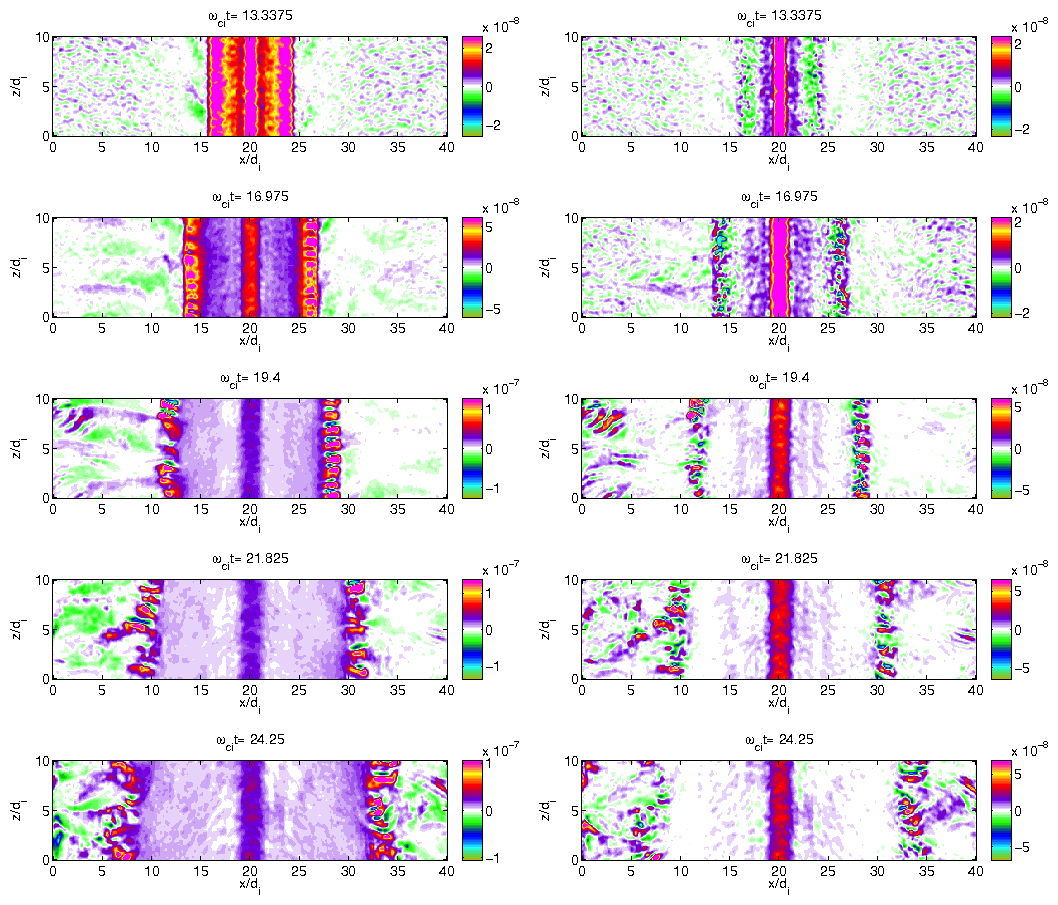}
\caption{\label{np-energy}False color representation of the energy exchange term $\bfJ\cdot\bfE$ in the simulation frame (left) and in the  frame $\bfJ\cdot\bfE^\prime=\bfJ\cdot(\bfE+\bfu_{CM}\times \bfB)$ moving with the local center of mass speed $\bfu_{CM}=(\bfu_e m_e+\bfu_i m_i)/(m_i+m_e)  $ (right). The results are shown as a $xz$ projection of the  neutral plane defined  for each $x$, $z$ by the $y$ location where  $|B_x|$ reaches its minimum. Different subsequent times are shown marked by the labels. All quantities are in code units.}
\end{figure}

A first most obvious consequence of the front instability described above is the exchange of large amounts of energy.  Figure \ref{np-energy} reports the energy exchange between fields and particles. The energy exchange is not frame independent. The left panels report the analysis in the simulation frame and the right panels that in the local frame moving with the local center of mass (CM) speed. The latter measure has the advantage of removing the energy expended in merely pushing the fronts outward, a work done by the Maxwell stress tensor against the pressure tensor.

Considering  the energy exchange in the simulation frame, the fronts are overwhelmingly the dominant loci of energy exchange. However this energy is mostly expended in moving the fronts. When the local CM frame is considered, the main energy exchange  $\bfJ\cdot\bfE^\prime$ is observed at the central x-line. 

Even in the $\bfJ\cdot\bfE^\prime$ measure, the fronts remain host to intense energy exchanges. Following the ripples of the front, one can notice how energy exchange alternates from positive to negative alternating generator regions (where the particle energy is transferred to the fields) to load regions (where electromagnetic energy goes to the particles). Recently, observational evidence has accumulated supporting such strong and at the same time oscillating energy exchanges at the fronts~\cite{hamrin2011energy,angelopoulos2013electromagnetic}. Based on observations from the Cluster\cite{hamrin2011energy} and THEMIS \cite{angelopoulos2013electromagnetic} missions, $\bfJ\cdot\bfE$ has been estimated locally, observing large values in the fronts, with a tendency for load regions to be found near the reconnection site and for a more equal distribution of load and generator regions further downstream from reconnection \cite{hamrin2011energy}. This evidence is obviously in agreement with an interpretation based on the drift-interchange mode discussed here~\cite{lapenta2014electromagnetic}.

The energy exchange measured in the frame moving with the center of mass speed is essentially equivalent \cite{marty-review} to the dissipation measure recently proposed by Zenitani and collaborators~\cite{zenitani2011new}. The virtue of this measure is to identifiy true local dissipation from energy linked to simple bulk motions. Figure \ref{np-energy} identifies that the front themselves as well as many other site downstream of the fronts are regions of intense energy exchanges. Not only the region of the fronts (located in the range $x/d_i\in[5,8]$ and  $x/d_i\in[32,35]$ at the final time) but also further downstream intense energy exchanges are observed. The present simulation uses open boundary conditions and a relatively large (for 3D runs) box along $x$, allowing us to follow a significant distance. Future work will be needed on even larger boxes to discover how far downstream these perturbations reach. 

Are these regions of intense  $\bfJ\cdot\bfE^\prime$ secondary reconnection sites? And are they surrounded by secondary electron diffusion regions? The next section will show that  indeed this is the case with several new reconnection sites generated by the perturbations introduced by the front instability.

\section{Evidence for secondary reconnection downstream of  drift-interchange unstable reconnection outflows}

\begin{figure}[h]
\includegraphics[width=\columnwidth]{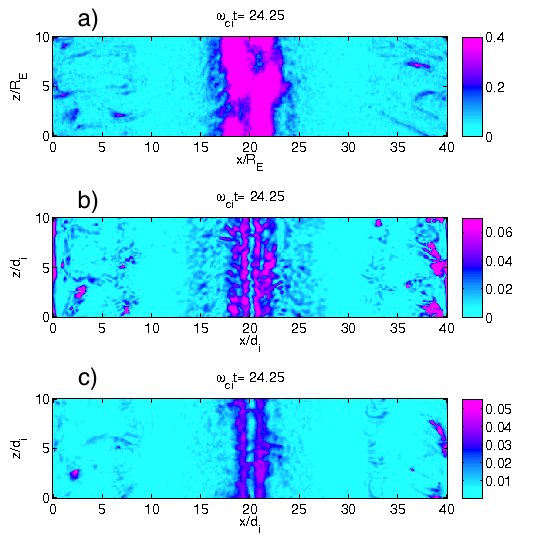}
\caption{\label{recon_measures}Three reconnection indicators on the neutral plane at the end of the run. From top to bottom: a) agyrotropy $A\O$ of the electron pressure tensor~\cite{scudder2008illuminating}; b) topological indicator $\left|\bfB\times \nabla \times \bfE_{||}\right|/B^2$, defined in \cite{hesse1988theoretical}; c) perpendicular velocity slippage, $\delta_\perp u_e/c= \left|\bfu_{e\perp} - \bfE\times\bfB/B^2\right|/c$.}
\end{figure}

Recently, Goldman et al. \cite{marty-review} analysed several measures of reconnection that identify topological changes and electron diffusion regions. We have recently applied such measures to show that secondary reconnection sites develop in the reconnection outflow~\cite{lapenta2015secondary}.

We extend here the analysis, by considering additional diagnostics (while referring the readers to the previous study for the others~\cite{lapenta2015secondary}) and considering a later time than earlier reported. Figure \ref{recon_measures} reports three among the several measures we analysed. All agree to give one clear answer: the regions of intense energy exchange identified above in the reconnection outflow are often (but not always) also reconnection regions. The three measures selected are:
\begin{itemize}
\item  The electron agyrotropy, defined via the two non zero eigenvalues of the auxiliary pressure tensor computed from the perpendicular velocities~\cite{scudder2008illuminating}:
$$
A\O=2\frac{\lambda_+-\lambda_-}
{\lambda_++\lambda_-}
$$
defined positive by definition, $\lambda_+>\lambda_-$.
Electron agyrotropy is taken to measure the extension of the EDR because agyrotropy leads to non ideal terms in the Ohm's law and violation of the frozen in condition.
\item The topological measure of violation of conservation of moving magnetic field lines $\left|\bfB\times \nabla \times \bfE_{||}\right|/B^2$ \cite{hesse1988theoretical}. A direct measure of reconnection in its topological consequences.
\item Perpendicular slippage, $\delta_\perp u_e/c= \left|\bfu_{e\perp} - \bfE\times\bfB/B^2\right|/c$, a direct measure that the electrons are not frozen in. 
\end{itemize}

As can be observed the same regions that were identified by the $\bfJ\cdot\bfE^\prime$ are in most cases also identified by the direct measures of reconnection and break of the frozen in condition. Recalling that we observed above the rippled fronts to be at $x/d_i\in[5,8]$ and  at $x/d_i\in[32,35]$ at the final time, the observed secondary reconnection sites are not just at the front but also further downstream. For example, prominent is a site at $x/d_i\approx 2.5, z/d_i\approx 2.5$, well downstream of the front.

\begin{figure}[h]
\includegraphics[width=\columnwidth]{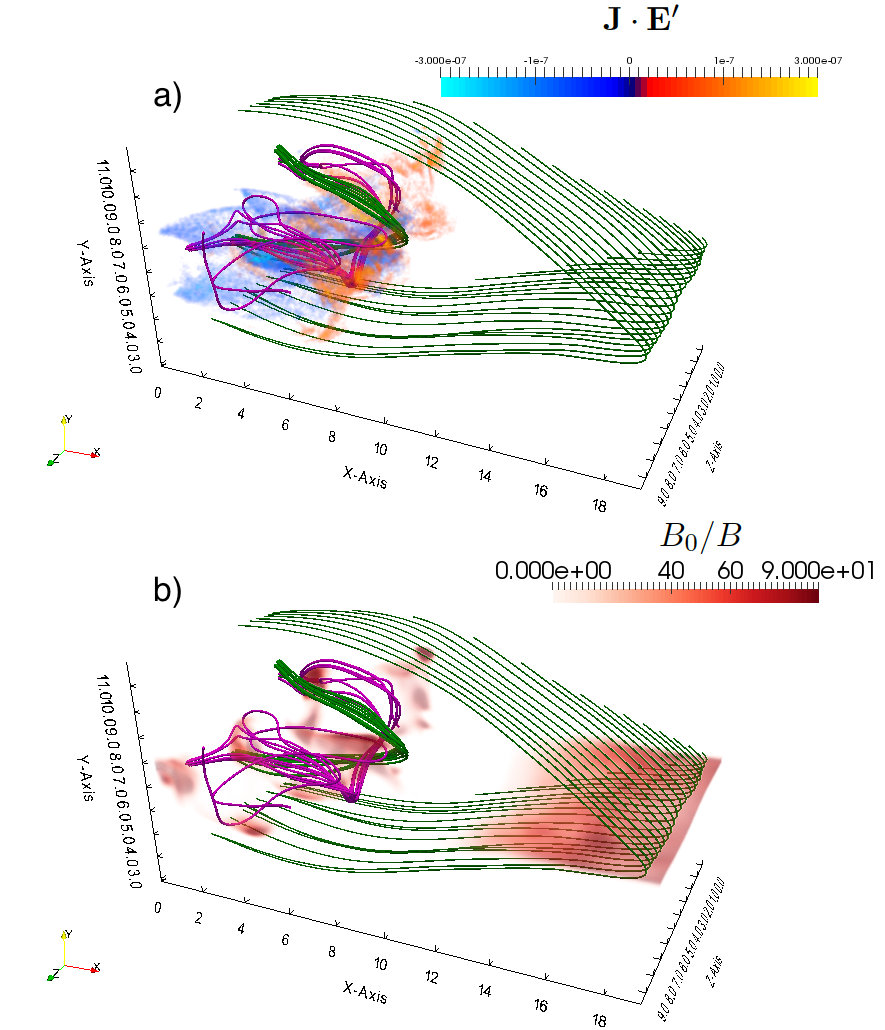}
\caption{\label{3drec}Three dimensional volume rendering of the energy exchange rate $\bfJ\cdot\bfE^\prime$  measure in the CM frame (a) and of the inverse of the magnetic field intensity $ B_0/B$  (b), a measure of local magnetic field vanishing. Selected field lines are show. A set of field lines just reconnected near the x-line is shown in green. Another set of field lines reconnected once at the x-line but already advected far down stream are shown also in green in proximity of a secondary reconnection site. A final set of field lines that have reconnected a second time at the secondary reconnection site is shown in violet.}
\end{figure}

The secondary reconnection sites are not limited to the neutral plane. At later times the current sheet becomes severely warped and its geometry is complex, see Fig.~\ref{bello}.  The analysis of reconnection indicators needs to be extended to full 3D visualization. 
Figure \ref{3drec} shows a 3D volume rendering  of half of the domain ($x<L_x/2$). Two quantities are shown, the energy exchange rate in  CM frame and the measure of smallness of the magnetic field intensity (defined as $B_0/B$). As can be observed the indicators are far from limited to a plane and become volume filling occupying a large potion of the region between the separatrices.

The ultimate proof of secondary reconnection is finding field lines whose topology is incompatible with reconnection happening only at the central x-line. Figure \ref{3drec} reports in green field lines reconnected only once by the x-line.  Two sets are shown. One freshly reconnected and still near the x-line and one already advected far downstream and located just before reaching a secondary reconnection site. 

 The newly reconnected field lines that have reconnected a second time in the secondary reconnection site are shown in violet. These field lines are not connected to the end wall in $x$ and are in fact spending their entire life inside the exhaust, meandering around. These lines are not initially there before reconnection start (see Ref. \cite{lapenta2015secondary} for a detailed analysis) and owe their existence to secondary reconnection.
 
\section{Conclusions}

In conclusion, we observe that the region of interaction between a reconnection exhaust and the surrounding environment becomes host to a drift-interchange instability that severely perturbs a large area of plasma downstream of the the interaction front. The area in between the separatrices becomes host of  secondary reconnection sites, each with its own EDR and topological changes. 

This finding is of great present interest for the MMS mission, increasing significantly the chances that EDR can be encountered by chance during the orbit of the four spacecraft. 
%
%

 \section*{Acknowledgments}
The present work is supported by  the NASA MMS Grant No. NNX08AO84G, by the Onderzoekfonds KU Leuven (Research Fund KU Leuven),
by the Interuniversity Attraction Poles Programme of the Belgian Science Policy
Office (IAP P7/08 CHARM)  and by the DEEP-ER project of the European Commission. The simulations were conducted on  NASA (NAS and NCCS) supercomputers, on  DOE-NERSC supercomputers, at the  VSC Flemish supercomputing centre and on the  facilities provided by 4 consecutive PRACE research infrastructure Tier-0 grants.

 \section*{References}
\providecommand{\newblock}{}

\end{document}